\documentclass[%
 aip,
 apl,
 amsmath,amssymb,
reprint,%
]{revtex4-1}

\usepackage{graphicx}
\usepackage{dcolumn}
\usepackage{bm}

\begin{document}

\title[Transient electrophoretic current]{Transient electrophoretic current in a nonpolar solvent}

\author{Pavel Kornilovitch}
 \email{pavel.kornilovich@hp.com}
 \affiliation{Hewlett-Packard Company, Imaging and Printing Division, Corvallis, Oregon 97330 USA} 
\author{Yoocharn Jeon}
 \email{yoocharn\_jeon@hp.com}
 \affiliation{Hewlett-Packard Laboratories, 1501 Page Mill Road, Palo Alto, California 94304 USA}

\date{21 March 2011}

\begin{abstract}

The transient electric current of surfactants dissolved in a nonpolar solvent is investigated both
experimentally and theoretically in the parallel-plate geometry. Due to a low concentration 
of free charges the cell can be completely polarized by an external voltage of several volts.
In this state, all the charged micelles are compacted against the electrodes. After the voltage 
is set to zero the reverse current features a sharp discharge spike and a broad peak. This shape 
and its variation with the compacting voltage are reproduced in a one-dimensional drift-diffusion 
model. The model reveals the broad peak is formed by a competition between an increasing number 
of charges drifting back to the middle of the cell and a decreasing electric field that drives 
the motion. After complete polarization is achieved, the shape of the peak stops evolving with 
further increase of the compacting voltage. The spike-peak separation time grows logarithmically 
with the charge content in the bulk. The time peak is a useful measure of the micelle mobility. 
Time integration of the peak yields the total charge in the system. By measuring its variation 
with temperature, the activation energy of bulk charge generation has been found to be 0.126 eV. 

\end{abstract}


\maketitle

\section{\label{sec:one}
Introduction
}

Electrical conduction in nonpolar fluids is of interest in relation to electrophoretic 
displays that use nonpolar-based solutions as the working 
medium.~\cite{Muerau1978,Novotny1979b,Jacobson1998}  When an appropriate surfactant is added 
to a nonpolar solvent its molecules form inverse micelles: globular aggregates with 
hydrophobic surfaces and hydrophilic interiors.~\cite{Eicke1980}  The micelles are believed 
to be responsible for the solution's nonzero electrical conductivity by stabilizing 
positive and negative charges and keeping them from recombining.~\cite{Morrison1993} 
In addition, the micelles have been suggested to charge colloidal particles by ionizing 
their surface groups and preferentially adsorbing on the surfaces.~\cite{Roberts2008} 

A useful way to gain insight into the microscopic properties of inverse micelles is through  
a detailed analysis of the transient current 
curves.~\cite{Novotny1979,Denat1982b,Novotny1986,Dikarev1997,Bazant2004a,Kim2005a,Bert2006a,Beunis2007a,Beunis2007b,Beunis2008,Prieve2008}
Typically, in such an experiment a single-surfactant solution is placed between two parallel 
electrodes, although more complex geometries have been used.~\cite{Kim2005a}  The voltage is 
stepped from zero to a finite value and then either back to zero or to some other value. 
The current recorded in the external circuit is an integral measure of the complex movement 
of charges inside the cell. By assuming the solution to be a symmetric electrolyte, the 
concentration of charged micelles, their mobility, diffusion coefficient, and the Stokes 
diameter can be inferred.     

Most of the previous work has focused on the sometimes very detailed quantitative analysis 
of the forward transient current.~\cite{Bazant2004a,Beunis2008}  However, in the early studies
it was noticed that the {\em reverse} transients often exhibit a nonmonotonic 
behavior.~\cite{Novotny1979b,Novotny1986}  Although the reverse peak has been observed in 
subsequent studies,~\cite{Bert2005,Bert2006a,Beunis2007b} it has not been analyzed or 
explained in sufficient detail. The present paper aims to contribute to this area.

\section{\label{sec:two}
Experiment
}

Poly-isobuthylene succinimide (OLOA 11000, Shevron Oronite Company LLC, 72\% active component) 
was dissolved in the hydrocarbon fluid Isopar\textsuperscript{\textregistered} M (Exxon Mobil 
Chemical, dielectric constant $\varepsilon = 2.0$) by sonication at different concentrations. 
No attempt was made to dry the components and the water content was not controlled. 
Parallel-plate cells were made from pieces of polyethylene terephthalate (PET) coated with conducting 
indium tin oxide (Sheldahl, Inc.). One side of the cell was photo-patterned with a square grid of 
SU8 spacer walls whereas the other side remained flat. The width of the walls was 10 $\mu$m and 
the grid period was 230 $\mu$m. The height of the walls (10 or 30 $\mu$m) defined the cell 
thickness. The active device area was about 10 cm$^2$. The voltage profile was provided by a DC 
power supply (Agilent Technologies). The current was recorded by a low-noise current preamplifier 
(Stanford Research Systems) at a sampling rate of 1000 Hz. Both instruments were controlled by a 
customized LabView application (National Instruments Corp.). To suppress the Faradaic processes, 
the measurements were done at below-room temperatures of $5^{\circ}$C, $10^{\circ}$C, and 
$15^{\circ}$C. The temperature was controlled by a convective environmental chamber 
(Autronic-Melchers GmbH) with accuracy of $0.1^{\circ}$C.

\begin{figure*}[t]
\includegraphics[width=0.98\textwidth]{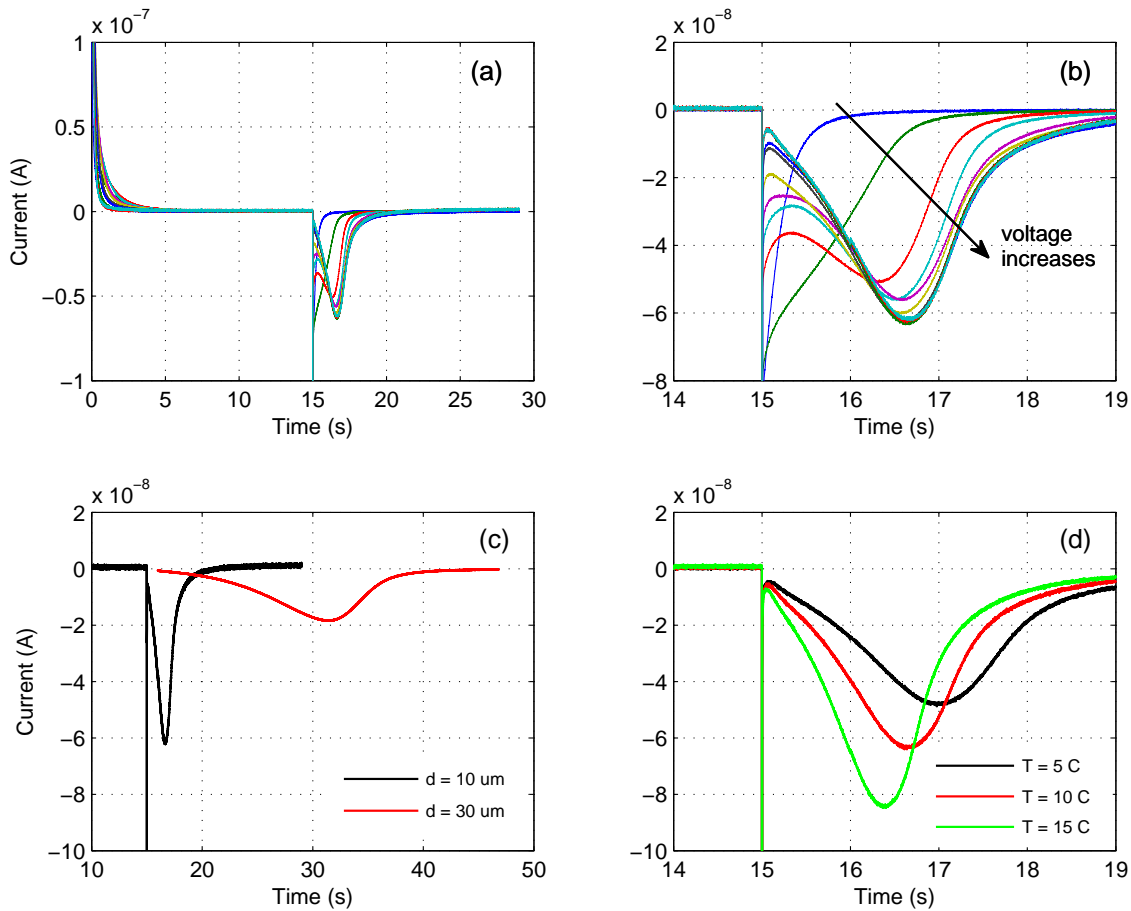}
\vspace{-0.5cm}
\caption{Transient currents of poly-isobuthylene succinimide in Isopar M in a parallel-plate 
cell. The voltage is stepped from zero to $V$ at 0 seconds and then back to zero at 15 seconds. 
(a) The overall shape of the transient for 0.5 wt \%. The cell gap is $d = 10$ $\mu$m and the 
temperature is $T = 10^{\circ}$C. 
The step voltages are 0.25, 0.5, 0.75, 1, 1.5, 2, 3, 4, 6, 8, and 10 V.
(b) A magnified view of the reverse peak region. Note the saturation of the shape after 3 V.  
(c) The cell gap dependence for 0.5 wt \%.  $V = 10$ volts.
(d) The $V = 8.0$ volts peaks for three temperatures of $5^{\circ}$C, $10^{\circ}$C, and 
$15^{\circ}$C. }   
\label{fig:two}
\end{figure*}

\section{\label{sec:three}
Observations
}

Transient currents of a 10 $\mu{\rm m}$-thick 0.5 wt.\% OLOA-11000 solution in Isopar M are 
shown in Fig.~\ref{fig:two}(a). The voltage between two parallel electrodes was stepped to a 
nonzero value between 0.25 and 10.0 V for 15 s, and then brought back to 0 V 
for another 14 seconds. As observed by other 
investigators~\cite{Novotny1979,Novotny1986,Bert2005,Bert2006a,Beunis2007b}
the currents possess the following characteristic features: a fast decay in the forward direction, 
then a sharp spike followed by a broad peak in the reverse direction. Initially, the forward decay 
is fast but then it slows down. Phenomenologically, the forward transient curve is well described
by an $RC$ circuit with a decreasing number of charge carriers (increasing $R$). This issue is not 
considered in the present work. The forward current continues to evolve with increasing voltage: 
it grows larger in magnitude and shorter in duration. In contrast, the reverse current quickly 
saturates with voltage. The $V = 2.0$ and 10.0 V curves are nearly identical. 
Figure~\ref{fig:two}(b) shows a magnified view of the reverse peak area. The peak develops between 
0.5 and 0.75 V and saturates by $V = 3$ V. This suggests 3 V cause complete polarization of the cell: 
all available charges are compacted near the electrodes and the charge cloud no longer changes with 
a further increase in voltage.      

Variation of the transient current with the cell thickness has been investigated. The reverse 
current for two thicknesses 10 $\mu$m and 30 $\mu$m are displayed in Fig.~\ref{fig:two}(c). As
expected intuitively, in the thicker cell, the reverse peak develops more slowly. The ratio of the peak 
times, $16.5/1.65 = 10.0$, is close to the ratio of the cell thicknesses squared, $3.0^2 = 9.0$. 
This scaling is discussed in the following sections. Variation of the current with temperature has 
also been investigated. Figure~\ref{fig:two}(d) compares the reverse transients for the three 
temperatures of $5^{\circ}$C, $10^{\circ}$C, and $15^{\circ}$C. As the temperature decreases the 
peak shifts to longer times and reduces in height.

\section{\label{sec:four}
Model
}

To develop an understanding of the observed behavior, a one-dimensional drift-diffusion 
(Nernst-Planck-Poisson) model has been applied. The model is formulated here in a dimensionless 
form.~\cite{Bazant2004a} Given the distance between two parallel electrodes $d$, the diffusion 
coefficient of {\em charged} micelles $D$ (both positive and negative), and the physical 
distance $x$ and time $t$, the dimensionless distance and time are introduced as $s = x/d$ 
and $\tau = (D/d^2) \, t$.  The electrostatic potential $\phi$ is measured in 
units of the thermal voltage: $\psi = (e/k_B T) \, \phi$.  The micelles are assumed to be 
single-charged. The concentration of micelles $n_{\pm}(x,t)$ is measured in units of the 
equilibrium initial concentration of one sign, $c_{\pm}(s,\tau) = n_{\pm}/n_0$.  The micelle 
mobility $\mu$ is related to the diffusion coefficient by the Einstein relation $D = k_B T \mu$. 
In such units, the micelle flux is given by a sum of the drift and diffusion contributions 
\begin{equation}
j_{\pm} = - \frac{\partial c_{\pm}}{\partial s} \mp c_{\pm} \frac{\partial \psi}{\partial s} \: .
\label{eq:one}
\end{equation}
In the simplest version of the model, the dissociation-recombination processes in the
bulk~\cite{Silver1965,Novotny1986,Strubbe2006} and the Fa\-ra\-da\-ic processes at the 
electrodes~\cite{Bockris1970,Novotny1986} are neglected. (Both are small for the experimental 
times and temperatures studied in this work.) As a result, the fluxes satisfy the local 
conservation laws  
\begin{equation}
\frac{\partial c_{\pm}}{\partial \tau} + \frac{\partial j_{\pm}}{\partial s} = 0 \: ,
\label{eq:two}
\end{equation}
and the blocking boundary conditions 
\begin{equation}
j_{\pm}(0,\tau) = j_{\pm}(1,\tau) = 0 \: . 
\label{eq:three}
\end{equation}
These conditions also imply global conservation of positive and negative charges. 
The electrostatic potential is governed by the Poisson equation
\begin{equation}
\frac{\partial^2 \psi}{\partial s^2} = - K \left( c_{+} - c_{-} \right) \: ,
\label{eq:four}
\end{equation}
and boundary conditions 
\begin{equation}
\psi(0,\tau) = 0 \, ; \hspace{0.7cm}
\psi(1,\tau) = \frac{e V(\tau)}{k_B T} \equiv u(\tau) \: . 
\label{eq:fivezero}
\end{equation}
The step-function boundary conditions are defined as
\begin{equation}
u(\tau) = 
\left\{ 
\begin{array}{lll}
0   , &             \tau   <  \tau_1      \\
u_0 , & \tau_1 \leq \tau \leq \tau_2      \\
0   , &             \tau   >  \tau_2 
\end{array} 
\right. \!\! .
\label{eq:five}
\end{equation}
In Eq.~(\ref{eq:four}), $K = e^2 d^2 n_0 / \varepsilon \varepsilon_0 k_B T$ is a dimensionless
parameter which is an equivalent measure of the total charge concentration. It can also 
be expressed as $K = d^2/2L^2_D$, where $L_D$ is the Debye length of the undisturbed solution. 
Another parameter is the magnitude of the voltage step $u_0$. The model is fully characterized by 
just two dimensionless parameters, $K$ and $u_0$. As such, it allows for complete numerical 
investigation. The initial conditions are uniform concentrations and a zero electrostatic potential:
\begin{equation}
c_{+} (s,0) = c_{-} (s,0) = 1 \: , \hspace{1.0cm} \psi(s,0) = 0 \: . 
\label{eq:six}
\end{equation}
Once $\psi(s,\tau)$ is known, the external current (in amperes) is computed as the time
derivative of the electric field at the electrode (as follows from the Gauss law):
\begin{equation}
I(\tau) = \frac{\varepsilon \varepsilon_0 A D k_B T}{e \, d^3}
\left. \frac{\partial^2 \psi(s,\tau)}{\partial s \: \partial \tau} \right\vert_{s = 0} \: 
\equiv I_0 \frac{\partial^2 \psi(0,\tau)}{\partial s \: \partial \tau} \: , 
\label{eq:seven}
\end{equation}
where $A$ is the electrode area.  

\begin{figure}[t]
\includegraphics[width=0.48\textwidth]{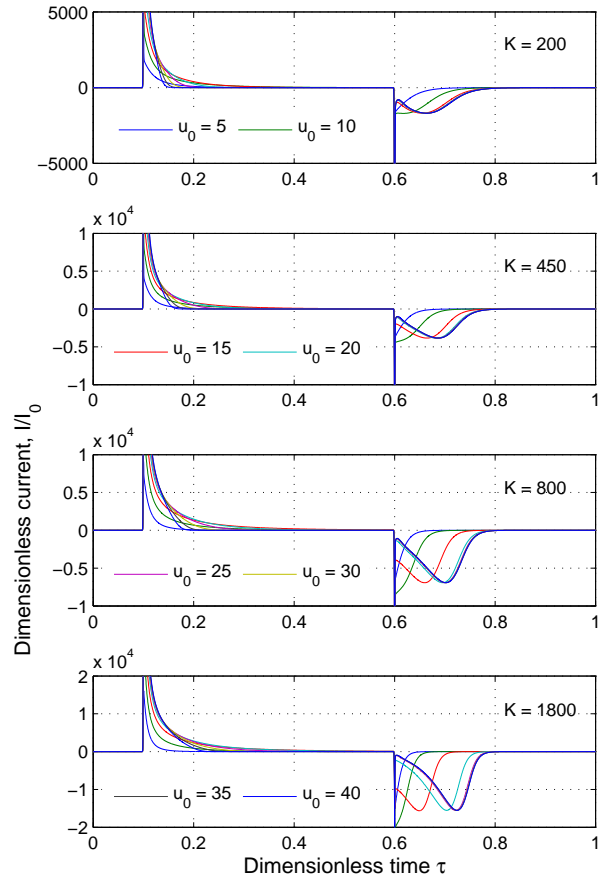}
\vspace{-0.5cm}
\caption{Numerical solution of equations (\ref{eq:one})-(\ref{eq:six}) for different values of 
$K$ and $u_0$. The voltage is stepped to $u_0$ at $\tau_1 = 0.1$ and back to zero at $\tau_2 = 0.6$. 
In all of the panels, $u_0$ increases from 5 to 40 in steps of 5. Notice the saturation of the 
reverse peak with $u_0$. Legend numbers apply to all panels.}
\label{fig:one}
\end{figure}

The Cauchy problem (\ref{eq:one})-(\ref{eq:six}) has been solved numerically using MATLAB (The 
MathWorks, Inc). The solution for several values of $K$ and $u_0$ is shown in Fig.~\ref{fig:one}. 
The model transient possesses the same qualitative features as the experimental one: a fast monotonic 
decay in the forward direction, then a sharp spike followed by a broad peak in the reverse direction. 
The reverse current saturates with $u_0$. The shape of the reverse peak slowly evolves with increasing 
$K$.

For the following discussion it is instructive to rewrite formula (\ref{eq:seven}) to make 
the relationship between the external current and the distribution of charges inside the 
cell more explicit. Integrating by parts and applying the Poisson equation (\ref{eq:four}) one obtains 
for the potential difference between the electrodes   
\begin{equation}
u(\tau) = \int^{1}_{0} \left( \frac{\partial \psi}{\partial s} \right) ds 
        = \frac{\partial \psi(1,\tau)}{\partial s} + K \int^{1}_{0} s \left( c_{+} - c_{-} \right) ds \: . 
\label{eq:nine}
\end{equation}
Note that due to global electroneutrality the electric fields near the two electrodes are 
always equal, or 
$[\partial \psi(s,\tau)/\partial s]_{s=0} = [\partial \psi(s,\tau)/\partial s]_{s=1}$. 
Substituting (\ref{eq:nine}) into (\ref{eq:seven}) one obtains
\begin{equation}
\frac{I(\tau)}{I_0} = \frac{d u(\tau)}{d\tau} 
- K \int^{1}_{0} s \left( \frac{\partial c_{+}(s,\tau)}{\partial \tau} 
                        - \frac{\partial c_{-}(s,\tau)}{\partial \tau} \right) ds \: .
\label{eq:ten}
\end{equation}
The dimensional form of the last formula reads
\begin{equation}
I(t) = \frac{\varepsilon \varepsilon_0 A}{d} \, \dot{V}(t) 
- \frac{e A}{d} \int^{d}_{0} x \left[ \dot{n}_{+}(x,t) - \dot{n}_{-}(x,t) \right] dx \: ,
\label{eq:eleven}
\end{equation}
where the dot means a partial derivative with respect to the real time $t$. The external 
current comprises two different contributions. The first term is charging and discharging of 
the geometric capacitor. (The coefficient by $\dot{V}$ is recognized as the geometric capacitance 
of the cell $C_g$.) The second term is an additional current associated with the movement of ions 
inside the cell. The first process is typically much faster than the second, and results in sharp 
charging/discharging initial spikes.  

Applying the conservation laws (\ref{eq:two}), integrating by parts, and making use of the
blocking boundary conditions (\ref{eq:three}), formula~(\ref{eq:ten}) can be rewritten as follows
\begin{equation}
\frac{I(\tau)}{I_0} = \frac{d u(\tau)}{d\tau} 
- K \int^{1}_{0} \left[ j_{+}(s,\tau) - j_{-}(s,\tau) \right] ds \: .
\label{eq:elevenone}
\end{equation}
Thus the long-term part of the external current is simply an integral of the ionic flux over the
cell thickness.

\begin{figure}[t]
\includegraphics[width=0.48\textwidth]{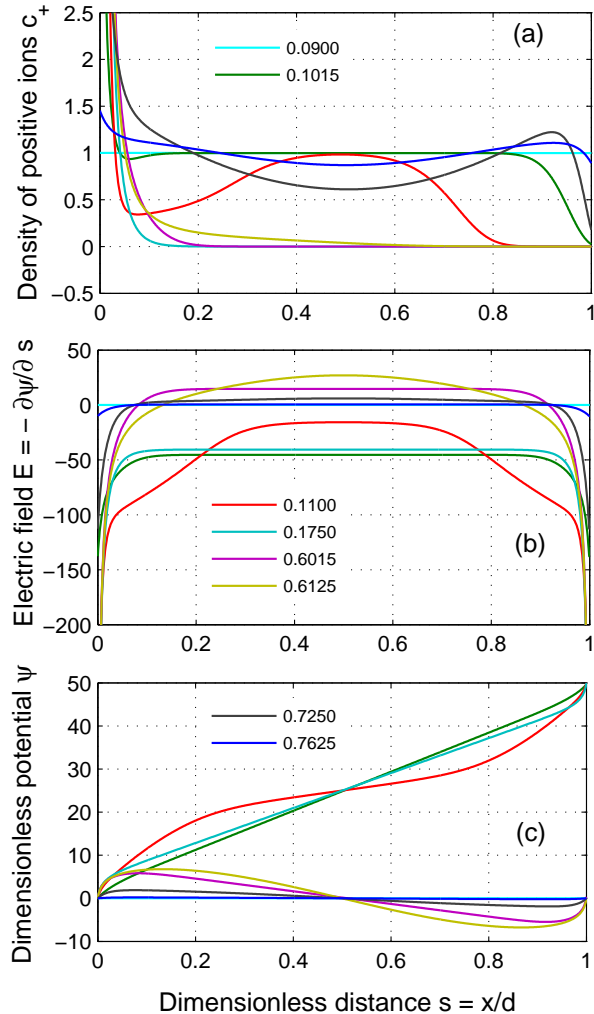}
\vspace{-0.5cm}
\caption{Spatial and temporal dependence of physical quantities for $K = 800$ and $u_0 = 50$. 
(a) Concentration of positive ions. (b) Electrostatic field. (c) Electrostatic potential. All 
quantities are dimensionless. The legend numbers are dimensionless times $\tau$. The legends are 
common for all three panels. The field is switched on at $\tau = 0.1$ and off at $\tau = 0.6$. 
The concentration of the negative ions is equal to that of the positive ions reflected around 
the cell center $s = 1/2$.}
\label{fig:five}
\end{figure}

\section{\label{sec:five}
Origin of the reverse peak
}

In addition to the external current, the numerical solution provides time-dependent 
distributions of the charges and the fields inside the cell, some of which are shown in 
Fig.~\ref{fig:five}. The origin of the reverse peak can be analyzed in sufficient detail, 
which is now described.   

During the forward phase the external electric field polarizes the cell by compacting the 
positive and negative ions against their respective electrodes. The equilibrium concentration 
profile is determined by the balance between the electrophoretic drift and diffusion when it 
reaches an equilibrium, cf. Fig.~\ref{fig:five}(a), $\tau = 0.1750$. As the field is turned off, 
the electrodes are brought to the same potential on a time scale much shorter than the 
characteristic diffusion time of the ions. A new potential distribution is established almost 
instantaneously after the switching at $\tau = 0.6$, cf. Fig.~\ref{fig:five}(c), $\tau = 0.6015$. 
This fast geometric capacitance discharge produces the sharp initial spike. 

Just after the switching most ions are still 
being pushed against the electrodes by the local electric field. However, the electrostatic 
potential of both electrodes is now zero. By symmetry, the potential in the cell center is also 
zero. (In a more general case of the asymmetric electrolyte or cell, the zero-potential point
will not be in the center of the cell. This fact does not change the following arguments.)
Therefore the potential assumes a local maximum somewhere near the negative electrode (and a 
local minimum near the positive electrode). This zero-field point must be within the charge cloud 
since by Poisson equation only uncompensated charge can change the curvature of the potential. 
As a result, to the right of the maximum point there is some positive charge, which is of the 
order of the geometric capacitance times the minimal voltage of total polarization. (The estimate
follows from the Gauss law.) Since the electric field is also positive there, the charge-field 
product creates a flux of positive ions toward the center of the cell. A similar drift of the 
negative ions away from the positive electrode is taking place at the opposite side of the cell. 
Closer to the electrodes, the drift and diffusion fluxes continue to effectively compensate each 
other.   

Initially, the integral ionic flux is small and, according to Eq.~(\ref{eq:elevenone}), the 
external current is also small. However, the external current is a flow of the electrode 
charge, and as a result the latter decreases. The reduction of the electrode charge moves the 
zero-field point deeper into the charge cloud thereby releasing more ions for drift-diffusion 
toward the bulk. The electric field increases inside the cell for the same reason. The integral 
ionic flux grows and so does the external current. Thus the current (i.e. time decrement of the 
electrode charge) is linearly related to the electrode charge itself. As a result, the current 
grows approximately {\em exponentially}, at least for short times when the positive and negative 
charge clouds do not overlap. As the clouds expand, positive and negative ions begin to mix in 
the middle of the cell reducing the space-charge and the electric field. The charge distribution 
becomes more uniform which dampens the internal fluxes and the external current. The entire 
process can be viewed as a competition between the increasing number of participating ions and 
decreasing driving fields. In the early stages of the reversal the increase in the ion number 
dominates, and the overall current increases with time. Later, the ion number levels off but the 
driving electric field subsides as does the current. 

The variation of the reverse current with the charge parameter $K \propto d^2 n_0$ is now 
discussed. As can be seen in Fig.~\ref{fig:one}, larger $K$s require larger voltages to reach 
current saturation. In addition, the peak maximum shifts to longer
times. The voltage increase can be understood from the following considerations. The reverse 
current saturates when the cell is fully polarized which occurs when $C_{dl} V > Q$. Here 
$C_{dl} \propto \sqrt{n_0} \, \exp{(u_p/4)}$ is the Gouy-Chapman double-layer capacitance and 
$Q \propto n_0 d$. At low charge densities, the polarization voltage scales as 
$u_p \propto d \sqrt{n_0} \propto \sqrt{K}$. At moderate and high densities the exponential 
enhancement of the capacitance is most important, and the scaling crosses over to 
$u_p \propto \ln (K)$. It is essential that the polarization voltage scales sublinearly with 
the density. The above estimates do not take into account the significant effects of the limited 
charge content in a finite cell on the double layer capacitance and polarization voltage. 
This issue was thoroughly analyzed by Verschueren et al.~\cite{Verschueren2008}

\begin{figure}[t]
\includegraphics[width=0.48\textwidth]{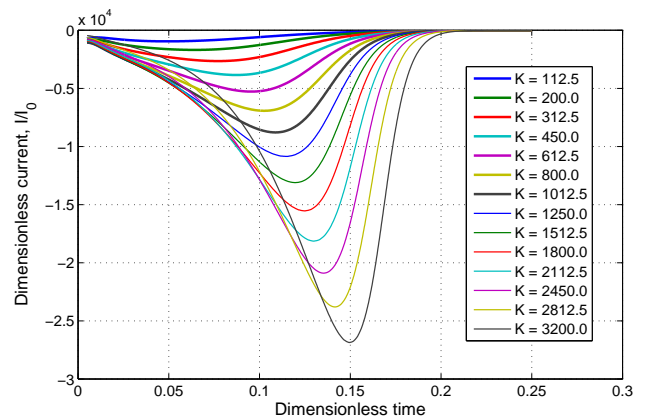}
\vspace{-0.5cm}
\caption{The saturated reverse current ($u_0 = 50$) at different values of the charge density 
parameter $K$. The reverse current is completely extinguished by $\tau = 0.2$. The curves have 
negative skewness whose magnitude increases with $K$. Notice a relatively weak dependence 
of the peak position on $K$.}
\label{fig:three}
\end{figure}

As mentioned above, the reverse drift starts with a small amount of charge and then accelerates 
exponentially by pulling ions from the cloud near the electrodes. The process continues until 
the charge ``stocks'' are depleted. The depletion time approximately corresponds to the peak 
time in the reverse current. Therefore the peak time should scale logarithmically with the 
total amount of charge in the system, $t_p \propto \ln (K)$. This scaling is confirmed by 
numerical calculations, as shown in Fig.~\ref{fig:three}. Notice the envelope of the curves' 
leading edges is roughly an exponential.

\section{\label{sec:six}
Micelle properties
}

The broad peak is the most conspicuous feature of the reverse transient current in the 
high-voltage regime. In this section it will be shown how useful physical information 
can be extracted from the time and shape of the peak. 

{\em Diffusivity.}
As can be seen in Fig.~\ref{fig:three}, the weak logarithmic dependence places the peak 
time within the interval $t_p = ( 0.10 \pm 0.05 ) ( d^2 / D )$ for a large spread of the
density parameter $K$ (a factor of 30). This suggests a simple method of estimating the ion's 
diffusivity: measure the reverse peak time $t_p$ in seconds, then $D \approx 0.1 ( d^2 / t_p )$ 
with the accuracy of 50\%. For example, for the data of Fig.~\ref{fig:two}(d), the peak times
are 2.00, 1.65, and 1.40 sec for $T = 5^{\circ}$C, $10^{\circ}$C, and $15^{\circ}$C. Using 
$d = 10.0$ $\mu{\rm m}$, one obtains the respective diffusivities of OLOA11000 in Isopar M 
at these temperatures as $D \approx 5.0$, 6.1, and 7.1 $\mu{\rm m}^2/{\rm sec}$. By the same
argument, the peak time should scale approximately quadratically with the cell thickness $d$,
cf. Fig.~\ref{fig:two}(c).  

\begin{figure}[t]
\includegraphics[width=0.48\textwidth]{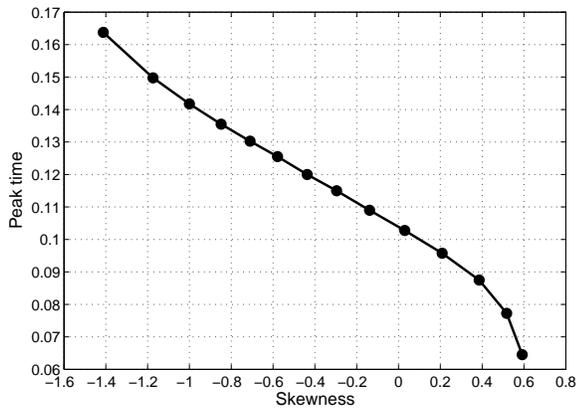}
\vspace{-0.5cm}
\caption{The dimensionless peak time of the saturated (i.e. at large $u_0$) reverse current 
as a function of the skewness. The time is counted from the moment the voltage is set to zero.}
\label{fig:four}
\end{figure}

A more accurate method involves taking into account the changes of the peak shape with $K$. 
A convenient measure of the current asymmetry is its skewness. Figure~\ref{fig:four} shows 
the peak time at saturation versus skewness, as calculated from the model data of 
Fig.~\ref{fig:three}. To avoid ambiguity in determining the skewness the data set is limited 
to a symmetric interval around the peak time. Once the skewness is calculated for an experimental 
reverse current the dimensionless peak time $\tau_p$ can be found in Fig.~\ref{fig:four}. 
Then the micelle diffusivity is $D = \tau_p (d^2/t_p)$. The skewness of the experimental data 
of Fig.~\ref{fig:two} are given in the fourth column of Table~\ref{tab:one}. The next column 
contains the respective dimensionless peak times $\tau_p$ determined from Fig.~\ref{fig:four}. 
They are 11-13 \% larger than the ``back-of-the-envelope'' value 0.1, which corrects the 
diffusivities to their final values given in the sixth column of the table.   

\begin{table*}
\renewcommand{\tabcolsep}{0.2cm}
\renewcommand{\arraystretch}{1.5}
\begin{tabular}{|c|c|c|c|c|c|c|c|c|c|c|}
\hline\hline
 $d$, $\mu{\rm m}$ & 
      $T$, $^{\circ}$C & 
                  $t_p$, s & 
                      Skewness & 
                          $\tau_p$ & 
                          $D$, m$^2$/s & 
                       $\mu_{\rm S}$, s/kg & 
                          $\eta$, mPa$\cdot$s  & 
                                   $a_{\rm S}$, nm & 
                                              $Q$, nC & 
                                               $n_0$, nmol/l \\ \hline
 10 &  5.0 & 2.00 & -0.28 & 0.115 & $5.8 \cdot 10^{-12}$ & $1.50 \cdot 10^{9}$ & 5.91 & 12.0 & 108.2 & 11.23 \\ \hline
 10 & 10.0 & 1.65 & -0.32 & 0.116 & $7.0 \cdot 10^{-12}$ & $1.80 \cdot 10^{9}$ & 4.81 & 12.3 & 117.0 & 12.15 \\ \hline
 10 & 15.0 & 1.40 & -0.38 & 0.118 & $8.4 \cdot 10^{-12}$ & $2.12 \cdot 10^{9}$ & 4.07 & 12.3 & 129.9 & 13.49 \\ \hline\hline
\end{tabular}
\caption{
Properties of inverse micelles of OLOA11000 in Isopar M extracted from the reverse transient peaks. 
$d$ is the cell thickness. $T$ is the experiment temperature. $t_p$ is the time of the reverse 
current peak counted from the voltage turn off. The fourth column is the skewness calculated from 
the data interval symmetric around the peak time. $D$ is the diffusion coefficient of the micelles.
$\mu_S$ is the Stokes mobility (the electrophoretic mobility divided by the elementary charge). 
$\eta$ is the viscosity of Isopar M. $a_S$ is the micelle Stokes diameter determined from 
Eq.~(\ref{eq:twelve}). $Q$ is the total charge of one sign in the cell determined from peak integration. 
$n_0$ is the volumetric concentration of charges micelles, in nano-mol/liter.}
\label{tab:one}
\end{table*}

{\em Micelles Stokes mobility and Stokes diameter.}
Once the diffusivity is known the Stokes mobility $\mu_S = D/k_B T_a$ and Stokes diameter 
\begin{equation}
a_S = \frac{1}{3 \pi \eta \, \mu_S} = \frac{k_B T_a}{3 \pi \eta \, D} \: ,
\label{eq:twelve}
\end{equation}
can be easily determined. Here $T_a$ is the absolute temperature in K, and $\eta$ is the viscosity of
the host fluid. The viscosity of Isopar M has been measured for 
$-25^{\circ}{\rm C} < T < +50^{\circ}{\rm C}$. 
Note that the Stokes mobility differs from the more familiar electrophoretic mobility by the 
elementary charge factor, $\mu_{ep} = e \, \mu_S$. The values of $\mu_S$, $a_S$ and $\eta$ 
for three temperatures are given in Table~\ref{tab:one}. Both $\mu_S$ and $\eta$ show significant
temperature dependence while the Stokes diameter does not. This suggests that the bulk of the 
temperature variation in the micelle mobility is due to the corresponding change of the viscosity
of the host fluid. A similar conclusion was reached by Novotny.~\cite{Novotny1986}  The overall 
estimate of the micelle diameter is $12.2(5)$ nm.       

A similar analysis of the $d = 30$ $\mu{\rm m}$ sample at $T = 10^{\circ}$C resulted in a slightly
larger mobility of $1.87 \cdot 10^{-12}$ s/kg and a slightly smaller micelle diameter of 11.8 nm. 
Note that the reverse transient current, cf. Fig.~\ref{fig:two}(c), is more skewed than the 
$d = 10$ $\mu{\rm m}$ curves. The corresponding skewness parameter, -0.76, is about twice as large.

{\em Equilibrium charge density.}
The saturated reverse transient curve allows an independent determination of the total amount of 
charge in the system. The Faradaic processes are suppressed at low temperatures and the bulk charge 
generation is too slow to affect the overall charge dynamics on the time scale of 
seconds.~\cite{Strubbe2006} Under these assumptions, the total amount of the charge of each polarity is 
conserved. This can be determined from the experimental transient curves. The argument is based on 
Eq.~(\ref{eq:eleven}). By integrating the second term from the moment just after the voltage turn 
off $t_1$ (i.e., just after the discharge spike) to infinity, one obtains 
\begin{equation}
\int^{\infty}_{t_1} I (t) \: dt = \frac{e \, A}{d} 
\int^{d}_{0} x \left[ n_{+}(x,t_1) - n_{-}(x,t_1) \right] dx  \: .
\label{eq:thirteen}
\end{equation}
The $t = \infty$ contribution vanishes since $n_{+}(x,\infty) = n_{-}(x,\infty) = n_0$, the micelles 
are fully mixed, and the integrand is identically zero. The volumetric densities $n_{\pm}(x,t_1)$ 
are those just after the voltage turn off. {\em In the ideal case of complete compaction}, the 
densities are delta functions: 
$n_{+}(x,t_1) = (n_0 d) \, \delta(x)$, $n_{-}(x,t_1) = (n_0 d) \, \delta(x-d)$. Performing the 
integration, the right hand side of Eq.~(\ref{eq:thirteen}) becomes $- e (A d) n_0$. Since $A d$ 
is the cell volume, the last expression is the total negative charge of the cell. In practice, the 
micelles are never compacted into a delta function; one can only speak about different degrees of 
approaching the limit. (In the experiment presented here, the compaction is believed to be almost 
complete, within a few percent of the limit.) The current integral provides the lower limit for the 
total charge:
\begin{equation}
\left\vert Q_{+} \right\vert = \left\vert Q_{-} \right\vert >  
\left\vert \int^{\infty}_{t_1} I (t) \: dt \right\vert  \: .
\label{eq:fourteen}
\end{equation}
Once $Q_{\pm}$ are known, the equilibrium concentrations of charged micelles are found by dividing 
by the cell volume. The values of $Q$ and $n_0$ are given in the last two columns of 
Table~\ref{tab:one}. The order of magnitude is $n_0 \sim 10^{-8}$ mol/l $\sim 10^{-5}$ mol/m$^3$, 
which is similar to the values reported in the literature.~\cite{Kim2005a}

{\em Activation energy.}
The temperature dependence of the equilibrium charge density allows us to estimate the activation 
energy $E_a$ of the charge generation process, which is of fundamental importance in dielectric 
fluids. Fitting the $n_0(T)$ values from Table~\ref{tab:one}, the activation energy is determined 
as $E_a$ = 0.126 eV = 2.91 kcal/mol. This value is close to the one reported by Novotny 
\cite{Novotny1986} for Aerosol OT surfactant (0.140 eV), which suggest that the ionization 
mechanisms in the two systems might be similar.

\section{ \label{sec:seven}
Summary
}

In summary, transient currents in a weakly conducting solution of poly-isobuthylene 
succinimide in Isopar M have been investigated. The limited charge content allows the complete 
polarization of the cell at low and medium voltages. Under these conditions, the bulk of 
the cell is devoid of mobile charges since bulk generation is a relatively slow process. 
The ability to control the bulk charge density makes this system a unique physical laboratory. 

The spike-peak structure is the robust feature of the experimental and model transient 
currents. The development of the peak signals the onset of cell polarization. The reverse 
current is nonmonotonic because even after the voltage turn off, the electrodes carry a
significant amount of charge. Initially, the latter keeps a significant portion of the 
mobile charges inside the cell close to the electrodes. As a result, the internal charges 
are released into the bulk not all at once, but gradually. At high external voltages  
the peak shape saturates. All of these features can be reproduced in a simple one-dimensional
drift-diffusion model. Such a model has been formulated in dimensionless units and     
solved numerically for a range of the two model parameters. The dimensionless peak time 
increases logarithmically with the charge density and lies in the interval $0.10 \pm 0.05$ 
for a wide range of parameters. Measured in real seconds, the peak time scales quadratically 
with the cell gap and inversely with the micelle mobility. The scaling provides a simple 
way to estimate the micelle diffusivity as $D \approx 0.1 \, d^2/t_p$. Taking into account 
the skewness of the peak allows a more accurate determination of the micelle parameters, 
which have been summarized in Table~\ref{tab:one}. The Stokes diameter of the micelles is 
estimated to be $12.2 \pm 0.5$ nm. From the temperature dependence of the total charge 
density, the activation energy of 0.126 eV has been inferred.

\begin{acknowledgments}

The authors wish to thank Gregg Combs for providing the current measurement hardware and 
software, Casey Lohrman for viscosity measurements, Tim Koch, Richard Henze and Kenneth Abbott
for support and encouragement, and all the colleagues at Hewlett-Packard for numerous 
discussions on the subject of this paper.  

\end{acknowledgments}


\begin{thebibliography}{10}%
\makeatletter
\providecommand \@ifxundefined [1]{%
 \ifx #1\undefined \expandafter \@firstoftwo
 \else \expandafter \@secondoftwo
\fi
}%
\providecommand \@ifnum [1]{%
 \ifnum #1\expandafter \@firstoftwo
 \else \expandafter \@secondoftwo
\fi
}%
\providecommand \enquote [1]{``#1''}%
\providecommand \bibnamefont  [1]{#1}%
\providecommand \bibfnamefont [1]{#1}%
\providecommand \citenamefont [1]{#1}%
\providecommand\href[0]{\@sanitize\@href}%
\providecommand\@href[1]{\endgroup\@@startlink{#1}\endgroup\@@href}%
\providecommand\@@href[1]{#1\@@endlink}%
\providecommand \@sanitize [0]{\begingroup\catcode`\&12\catcode`\#12\relax}%
\@ifxundefined \pdfoutput {\@firstoftwo}{%
 \@ifnum{\z@=\pdfoutput}{\@firstoftwo}{\@secondoftwo}%
}{%
 \providecommand\@@startlink[1]{\leavevmode}%
 \providecommand\@@endlink[0]{}%
}{%
 \providecommand\@@startlink[1]{%
  \leavevmode
  \pdfstartlink
   attr{/Border[0 0 1 ]/H/I/C[0 1 1]}%
   user{/Subtype/Link/A<</Type/Action/S/URI/URI(#1)>>}%
  \relax
 }%
 \providecommand\@@endlink[0]{\pdfendlink}%
}%
\providecommand \url  [0]{\begingroup\@sanitize \@url }%
\providecommand \@url [1]{\endgroup\@href {#1}{\urlprefix}}%
\providecommand \urlprefix [0]{URL }%
\providecommand \Eprint[0]{\href }%
\@ifxundefined \urlstyle {%
  \providecommand \doi [1]{doi:\discretionary{}{}{}#1}%
}{%
  \providecommand \doi [0]{doi:\discretionary{}{}{}\begingroup
  \urlstyle{rm}\Url }%
}%
\providecommand \doibase [0]{http://dx.doi.org/}%
\providecommand \Doi[1]{\href{\doibase#1}}%
\providecommand \selectlanguage [0]{\@gobble}%
\providecommand \bibinfo [0]{\@secondoftwo}%
\providecommand \bibfield [0]{\@secondoftwo}%
\providecommand \translation [1]{[#1]}%
\providecommand \BibitemOpen[0]{}%
\providecommand \bibitemStop [0]{}%
\providecommand \bibitemNoStop [0]{.\EOS\space}%
\providecommand \EOS [0]{\spacefactor3000\relax}%
\providecommand \BibitemShut [1]{\csname bibitem#1\endcsname}%
\bibitem{Muerau1978}%
  \BibitemOpen
  \bibfield{author}{%
  \bibinfo {author} {\bibfnamefont{P.}~\bibnamefont{M{\"u}rau}}\ and\ \bibinfo
  {author} {\bibfnamefont{B.}~\bibnamefont{Singer}},\ }%
  \bibfield{journal}{%
  \bibinfo {journal} {J. Appl. Phys.}\ }%
  \textbf{\bibinfo {volume} {49}},\ \bibinfo {pages} {4820} (\bibinfo {year}
  {1978})\BibitemShut{NoStop}%
\bibitem{Novotny1979b}%
  \BibitemOpen
  \bibfield{author}{%
  \bibinfo {author} {\bibfnamefont{V.}~\bibnamefont{Novotny}}\ and\ \bibinfo
  {author} {\bibfnamefont{M.~A.}\ \bibnamefont{Hopper}},\ }%
  \bibfield{journal}{%
  \bibinfo {journal} {J. Electrochem. Soc.}\ }%
  \textbf{\bibinfo {volume} {126}},\ \bibinfo {pages} {2211} (\bibinfo {year}
  {1979})\BibitemShut{NoStop}%
\bibitem{Jacobson1998}%
  \BibitemOpen
  \bibfield{author}{%
  \bibinfo {author} {\bibfnamefont{B.}~\bibnamefont{Comiskey}}, \bibinfo
  {author} {\bibfnamefont{J.~D.}\ \bibnamefont{Albert}}, \bibinfo {author}
  {\bibfnamefont{H.}~\bibnamefont{Yoshizawa}},\ and\ \bibinfo {author}
  {\bibfnamefont{J.}~\bibnamefont{Jacobson}},\ }%
  \bibfield{journal}{%
  \bibinfo {journal} {Nature}\ }%
  \textbf{\bibinfo {volume} {394}},\ \bibinfo {pages} {253} (\bibinfo {year}
  {1998})\BibitemShut{NoStop}%
\bibitem{Eicke1980}%
  \BibitemOpen
  \bibfield{author}{%
  \bibinfo {author} {\bibfnamefont{H.-F.}\ \bibnamefont{Eicke}},\ }%
  \enquote{\bibinfo {title} {Micelles},}\ \ (\bibinfo {publisher} {Springer},\
  \bibinfo {address} {Berlin, Heidelberg},\ \bibinfo {year} {1980})\ Chap.\
  \bibinfo {chapter} {2, Surfactants in Nonpolar Solvents. Aggregation and
  Micellization}, pp.\ \bibinfo {pages} {85--145}\BibitemShut{NoStop}%
\bibitem{Morrison1993}%
  \BibitemOpen
  \bibfield{author}{%
  \bibinfo {author} {\bibfnamefont{I.~D.}\ \bibnamefont{Morrison}},\ }%
  \bibfield{journal}{%
  \bibinfo {journal} {Colloids and Surfaces A}\ }%
  \textbf{\bibinfo {volume} {71}},\ \bibinfo {pages} {1} (\bibinfo {year}
  {1993})\BibitemShut{NoStop}%
\bibitem{Roberts2008}%
  \BibitemOpen
  \bibfield{author}{%
  \bibinfo {author} {\bibfnamefont{G.~S.}\ \bibnamefont{Roberts}}, \bibinfo
  {author} {\bibfnamefont{R.}~\bibnamefont{Sanchez}}, \bibinfo {author}
  {\bibfnamefont{R.}~\bibnamefont{Kemp}}, \bibinfo {author}
  {\bibfnamefont{T.}~\bibnamefont{Wood}},\ and\ \bibinfo {author}
  {\bibfnamefont{P.}~\bibnamefont{Bartlett}},\ }%
  \bibfield{journal}{%
  \bibinfo {journal} {Langmuir}\ }%
  \textbf{\bibinfo {volume} {24}},\ \bibinfo {pages} {6530} (\bibinfo {year}
  {2008})\BibitemShut{NoStop}%
\bibitem{Novotny1979}%
  \BibitemOpen
  \bibfield{author}{%
  \bibinfo {author} {\bibfnamefont{V.}~\bibnamefont{Novotny}}\ and\ \bibinfo
  {author} {\bibfnamefont{M.~A.}\ \bibnamefont{Hopper}},\ }%
  \bibfield{journal}{%
  \bibinfo {journal} {J. Electrochem. Soc.}\ }%
  \textbf{\bibinfo {volume} {126}},\ \bibinfo {pages} {925} (\bibinfo {year}
  {1979})\BibitemShut{NoStop}%
\bibitem{Denat1982b}%
  \BibitemOpen
  \bibfield{author}{%
  \bibinfo {author} {\bibfnamefont{A.}~\bibnamefont{Denat}}, \bibinfo {author}
  {\bibfnamefont{B.}~\bibnamefont{Gosse}},\ and\ \bibinfo {author}
  {\bibfnamefont{J.~P.}\ \bibnamefont{Gosse}},\ }%
  \bibfield{journal}{%
  \bibinfo {journal} {J. Electrostatics}\ }%
  \textbf{\bibinfo {volume} {12}},\ \bibinfo {pages} {197} (\bibinfo {year}
  {1982})\BibitemShut{NoStop}%
\bibitem{Novotny1986}%
  \BibitemOpen
  \bibfield{author}{%
  \bibinfo {author} {\bibfnamefont{V.}~\bibnamefont{Novotny}},\ }%
  \bibfield{journal}{%
  \bibinfo {journal} {J. Electrochem. Soc.}\ }%
  \textbf{\bibinfo {volume} {133}},\ \bibinfo {pages} {1629} (\bibinfo {year}
  {1986})\BibitemShut{NoStop}%
\bibitem{Dikarev1997}%
  \BibitemOpen
  \bibfield{author}{%
  \bibinfo {author} {\bibfnamefont{B.~N.}\ \bibnamefont{Dikarev}}, \bibinfo
  {author} {\bibfnamefont{G.~G.}\ \bibnamefont{Karasev}}, \bibinfo {author}
  {\bibfnamefont{V.~I.}\ \bibnamefont{Bolshakov}}, \bibinfo {author}
  {\bibfnamefont{R.~G.}\ \bibnamefont{Romanets}},\ and\ \bibinfo {author}
  {\bibfnamefont{I.~V.}\ \bibnamefont{Potapov}},\ }%
  \bibfield{journal}{%
  \bibinfo {journal} {J. Electrostatics}\ }%
  \textbf{\bibinfo {volume} {40-41}},\ \bibinfo {pages} {147} (\bibinfo {year}
  {1997})\BibitemShut{NoStop}%
\bibitem{Bazant2004a}%
  \BibitemOpen
  \bibfield{author}{%
  \bibinfo {author} {\bibfnamefont{M.~Z.}\ \bibnamefont{Bazant}}, \bibinfo
  {author} {\bibfnamefont{K.}~\bibnamefont{Thornton}},\ and\ \bibinfo {author}
  {\bibfnamefont{A.}~\bibnamefont{Ajdari}},\ }%
  \bibfield{journal}{%
  \bibinfo {journal} {Phys. Rev. E}\ }%
  \textbf{\bibinfo {volume} {70}},\ \bibinfo {pages} {021506} (\bibinfo {year}
  {2004})\BibitemShut{NoStop}%
\bibitem{Kim2005a}%
  \BibitemOpen
  \bibfield{author}{%
  \bibinfo {author} {\bibfnamefont{J.}~\bibnamefont{Kim}}, \bibinfo {author}
  {\bibfnamefont{J.~L.}\ \bibnamefont{Anderson}}, \bibinfo {author}
  {\bibfnamefont{S.}~\bibnamefont{Garoff}},\ and\ \bibinfo {author}
  {\bibfnamefont{L.~J.~M.}\ \bibnamefont{Schlangen}},\ }%
  \bibfield{journal}{%
  \bibinfo {journal} {Langmuir}\ }%
  \textbf{\bibinfo {volume} {21}},\ \bibinfo {pages} {8620} (\bibinfo {year}
  {2005})\BibitemShut{NoStop}%
\bibitem{Bert2006a}%
  \BibitemOpen
  \bibfield{author}{%
  \bibinfo {author} {\bibfnamefont{T.}~\bibnamefont{Bert}}, \bibinfo {author}
  {\bibfnamefont{H.~D.}\ \bibnamefont{Smet}}, \bibinfo {author}
  {\bibfnamefont{F.}~\bibnamefont{Beunis}},\ and\ \bibinfo {author}
  {\bibfnamefont{K.}~\bibnamefont{Neyts}},\ }%
  \bibfield{journal}{%
  \bibinfo {journal} {Displays}\ }%
  \textbf{\bibinfo {volume} {27}},\ \bibinfo {pages} {50} (\bibinfo {year}
  {2006})\BibitemShut{NoStop}%
\bibitem{Beunis2007a}%
  \BibitemOpen
  \bibfield{author}{%
  \bibinfo {author} {\bibfnamefont{F.}~\bibnamefont{Beunis}}, \bibinfo {author}
  {\bibfnamefont{F.}~\bibnamefont{Strubbe}}, \bibinfo {author}
  {\bibfnamefont{K.}~\bibnamefont{Neyts}},\ and\ \bibinfo {author}
  {\bibfnamefont{A.~R.~M.}\ \bibnamefont{Verschueren}},\ }%
  \bibfield{journal}{%
  \bibinfo {journal} {Applied Physics Letters}\ }%
  \textbf{\bibinfo {volume} {90}},\ \bibinfo {pages} {182103} (\bibinfo {year}
  {2007})\BibitemShut{NoStop}%
\bibitem{Beunis2007b}%
  \BibitemOpen
  \bibfield{author}{%
  \bibinfo {author} {\bibfnamefont{F.}~\bibnamefont{Beunis}}, \bibinfo {author}
  {\bibfnamefont{F.}~\bibnamefont{Strubbe}}, \bibinfo {author}
  {\bibfnamefont{M.}~\bibnamefont{Marescaux}}, \bibinfo {author}
  {\bibfnamefont{K.}~\bibnamefont{Neyts}},\ and\ \bibinfo {author}
  {\bibfnamefont{A.~R.~M.}\ \bibnamefont{Verschueren}},\ }%
  \bibfield{journal}{%
  \bibinfo {journal} {Applied Physics Letters}\ }%
  \textbf{\bibinfo {volume} {91}},\ \bibinfo {pages} {182911} (\bibinfo {year}
  {2007})\BibitemShut{NoStop}%
\bibitem{Beunis2008}%
  \BibitemOpen
  \bibfield{author}{%
  \bibinfo {author} {\bibfnamefont{F.}~\bibnamefont{Beunis}}, \bibinfo {author}
  {\bibfnamefont{F.}~\bibnamefont{Strubbe}}, \bibinfo {author}
  {\bibfnamefont{M.}~\bibnamefont{Marescaux}}, \bibinfo {author}
  {\bibfnamefont{J.}~\bibnamefont{Beeckman}}, \bibinfo {author}
  {\bibfnamefont{K.}~\bibnamefont{Neyts}},\ and\ \bibinfo {author}
  {\bibfnamefont{A.~R.~M.}\ \bibnamefont{Verschueren}},\ }%
  \bibfield{journal}{%
  \bibinfo {journal} {Phys. Rev. E}\ }%
  \textbf{\bibinfo {volume} {78}},\ \bibinfo {pages} {011502} (\bibinfo {year}
  {2008})\BibitemShut{NoStop}%
\bibitem{Prieve2008}%
  \BibitemOpen
  \bibfield{author}{%
  \bibinfo {author} {\bibfnamefont{D.~C.}\ \bibnamefont{Prieve}}, \bibinfo
  {author} {\bibfnamefont{J.~D.}\ \bibnamefont{Hoggard}}, \bibinfo {author}
  {\bibfnamefont{R.}~\bibnamefont{Fu}}, \bibinfo {author}
  {\bibfnamefont{P.~J.}\ \bibnamefont{Sides}},\ and\ \bibinfo {author}
  {\bibfnamefont{R.}~\bibnamefont{Bethea}},\ }%
  \bibfield{journal}{%
  \bibinfo {journal} {Langmuir}\ }%
  \textbf{\bibinfo {volume} {24}},\ \bibinfo {pages} {1120} (\bibinfo {year}
  {2008})\BibitemShut{NoStop}%
\bibitem{Bert2005}%
  \BibitemOpen
  \bibfield{author}{%
  \bibinfo {author} {\bibfnamefont{T.}~\bibnamefont{Bert}}, \bibinfo {author}
  {\bibfnamefont{H.~D.}\ \bibnamefont{Smet}}, \bibinfo {author}
  {\bibfnamefont{F.}~\bibnamefont{Beunis}},\ and\ \bibinfo {author}
  {\bibfnamefont{F.}~\bibnamefont{Strubbe}},\ }%
  \bibfield{journal}{%
  \bibinfo {journal} {Opto-Electronics Review}\ }%
  \textbf{\bibinfo {volume} {13}},\ \bibinfo {pages} {281} (\bibinfo {year}
  {2005})\BibitemShut{NoStop}%
\bibitem{Silver1965}%
  \BibitemOpen
  \bibfield{author}{%
  \bibinfo {author} {\bibfnamefont{M.}~\bibnamefont{Silver}},\ }%
  \bibfield{journal}{%
  \bibinfo {journal} {J. Chem. Phys.}\ }%
  \textbf{\bibinfo {volume} {42}},\ \bibinfo {pages} {1011} (\bibinfo {year}
  {1965})\BibitemShut{NoStop}%
\bibitem{Strubbe2006}%
  \BibitemOpen
  \bibfield{author}{%
  \bibinfo {author} {\bibfnamefont{F.}~\bibnamefont{Strubbe}}, \bibinfo
  {author} {\bibfnamefont{A.~R.~M.}\ \bibnamefont{Verschueren}}, \bibinfo
  {author} {\bibfnamefont{L.~J.~M.}\ \bibnamefont{Schlangen}}, \bibinfo
  {author} {\bibfnamefont{F.}~\bibnamefont{Beunis}},\ and\ \bibinfo {author}
  {\bibfnamefont{K.}~\bibnamefont{Neyts}},\ }%
  \bibfield{journal}{%
  \bibinfo {journal} {Journal of Colloid and Interface Science}\ }%
  \textbf{\bibinfo {volume} {300}},\ \bibinfo {pages} {396} (\bibinfo {year}
  {2006})\BibitemShut{NoStop}%
\bibitem{Bockris1970}%
  \BibitemOpen
  \bibfield{author}{%
  \bibinfo {author} {\bibfnamefont{J.~O.}\ \bibnamefont{Bockris}}\ and\
  \bibinfo {author} {\bibfnamefont{A.~K.~N.}\ \bibnamefont{Reddy}},\ }%
  \emph{\bibinfo {title} {Modern Electrochemistry}},\ Vol.~\bibinfo {volume}
  {2}\ (\bibinfo {publisher} {Plenum},\ \bibinfo {address} {New York},\
  \bibinfo {year} {1970})\BibitemShut{NoStop}%
\bibitem{Verschueren2008}%
  \BibitemOpen
  \bibfield{author}{%
  \bibinfo {author} {\bibfnamefont{A.~R.~M.}\ \bibnamefont{Verschueren}}, 
  \bibinfo {author} {\bibfnamefont{P.~H.~L.}\ \bibnamefont{Notten}},
  \bibinfo {author} {\bibfnamefont{L.~J.~M.}\ \bibnamefont{Schlangen}},  
  \bibinfo {author} {\bibfnamefont{F.}~\bibnamefont{Strubbe}}, 
  \bibinfo {author} {\bibfnamefont{F.}~\bibnamefont{Beunis}},\ and\ 
  \bibinfo {author} {\bibfnamefont{K.}~\bibnamefont{Neyts}},\ }%
  \bibfield{journal}{%
  \bibinfo {journal} {Journal of Physical Chemistry B}\ }%
  \textbf{\bibinfo {volume} {112}},\ \bibinfo {pages} {13038} (\bibinfo {year}
  {2008})\BibitemShut{NoStop}%
\end{thebibliography}

\providecommand{\noopsort}[1]{}\providecommand{\singleletter}[1]{#1}%

\end{document}